\documentclass{article}

\usepackage{hyperref}
\usepackage{graphicx}
\usepackage{wrapfig}
 \usepackage{geometry}

    \newgeometry{vmargin={20mm}, hmargin={37mm,37mm}}   
\usepackage{authblk}
\newcommand{\hl}[1]{#1}

\begin{document}

\title{The Necessity of AI Audit Standards Boards}

\author[1,2]{David Manheim}
\affil[1]{Technion - Israel Institute of Technology}
\affil[2]{ALTER (Association for Long Term Existence and Resilience)}

\author[3]{Sammy Martin}
\affil[3]{Transformative Futures Institute}

\author[4]{Mark Bailey}
\affil[4]{National Intelligence University}

\author[5]{Mikhail Samin}
\affil[5]{Independent}

\author[3]{Ross Greutzmacher}
\date{}                     
\setcounter{Maxaffil}{0}
\renewcommand\Affilfont{\itshape\small}

\maketitle
\begin{abstract}
  Auditing of AI systems is a promising way to understand and manage
ethical problems and societal risks associated with contemporary AI
systems, as well as some anticipated future risks. Efforts to develop
standards for auditing Artificial Intelligence (AI) systems have
therefore understandably gained momentum. However, we argue that
creating auditing standards is not just insufficient, but actively
harmful by proliferating unheeded and inconsistent standards, especially
in light of the rapid evolution and ethical and safety challenges of AI.
Instead, the paper proposes the establishment of an AI Audit Standards
Board, responsible for developing and updating auditing methods and
standards in line with the evolving nature of AI technologies. Such a
body would ensure that auditing practices remain relevant, robust, and
responsive to the rapid advancements in AI. The paper argues that such a
governance structure would also be helpful for maintaining public trust
in AI and for promoting a culture of safety and ethical responsibility
within the AI industry.

Throughout the paper, we draw parallels with other industries, including
safety-critical industries like aviation and nuclear energy, as well as
more prosaic ones such as financial accounting and pharmaceuticals. AI
auditing should emulate those fields, and extend beyond technical
assessments to include ethical considerations and stakeholder
engagement, but we explain that this is not enough; emulating other
fields' governance mechanisms for these processes, and for audit
standards creation, is a necessity. We also emphasize the importance of
auditing the entire development process of AI systems, not just the
final products. The proposed approach will help ensure that auditing AI
does not devolve into safety washing, and instead directly addresses
risks and ethical concerns that will continue to arise as AI becomes
increasingly important in society, and as human interaction with these
systems changes over time.
\end{abstract}

\noindent
\textbf{CCS CONCEPTS:} Computing / technology policy {\small{}•} Human and societal
aspects of security and privacy\\

\begin{sloppypar}
\noindent \textbf{Additional Keywords and Phrases:} AI Governance, AI Audit,
Audit, Ethical AI, AI Policy, \mbox{Organizational Culture}, \mbox{Audit Standards}, Standards Setting
\end{sloppypar}
\maketitle

\section{Introduction}\label{introduction}

Audits are used in different domains both for giving an account of what
is happening within a system, and verification of requirements.
(Courville et al. 2003) When considering how to perform audits in a new
domain, drawing on best practices from other domains is critical. There
is now significant focus on audits for ensuring the safety and
evaluating the risks and harms of AI systems, (Mökander et al. 2023,
Shevlane et al. 2023; Sharkey et al. 2023) as well as significant
earlier work on audit methods for evaluation of societal implications
(Raji et al. 2020) and on what a mature ethics process involves.
(Krijger et al. 2022) While it is encouraging to see action addressing
the critical role of evaluating and auditing risks from frontier models
and on evaluation of ethical standards, there are many challenges for
these types of evaluations and audits. (Costanza-Chock et al 2022)
Below, we argue that the current approach of standards development for
AI systems is harmful, among other reasons, due to proliferating,
inconsistent, and rapidly outdated static standards, along with lack of
clarity about what is appropriate in any given domain, for any specific
AI system, or application. This fragments efforts, undermines efforts to
make any specific auditing methods standard, and reduces the usefulness
of standards development.

To supplement important technical approaches being developed for
auditing AI systems, we need a process for broader audits, and for
ongoing development of standards --- an audit standards body, not just
audit standards. To explain what is needed and why, we review past work
and provide a brief background on current audit approaches. We then
explain why static standards and current efforts are fundamentally
incapable of addressing the relevant ethical challenges and risks. We
also note that auditing standards are not the same thing as standards
for audits, and neither necessarily implies regulation. In addition,
different methods, audit approaches, and standards are needed for
different model types and applications. (Frase, 2023) This is especially
true when specific standards are unclear or disputed, or when detailed
standards would be unwise, as we will explore. Therefore, the body of
the paper revisits some known ideas in AI audit, as well as some
drawbacks of some of the approaches methods, both to provide a brief
overview of the issues that evaluation and auditing should ideally
address, and to show how the suggestion of audit boards differs from
other approaches to standards.

\subsection{Background}\label{background}

The idea of auditing IT systems is well established in the context of
computing. (van Biene-Hershey, 2007) Standards for doing so, such as
COBIT, date at least to the 1990s, significantly predating current
paradigms in artificial Intelligence. Recently, debate about safety,
misuse, and bias has led to internal and external checks on models,
notably including the internal evaluations and staged release of GPT-2.
(Shevlane, 2022) Continuing in the tradition of auditing, there have
been audits specific to AI systems, including both a growing ecosystem
of AI ethics and accountability audits, (Birhane et al 2014) as well as
safety audit efforts such as the red-teaming performed for GPT-4.
(OpenAI 2023a, 2023b)

Self-reporting via ``Model Cards'' (Mitchell et al 2019) has become
commonplace, and internal auditing and red-teaming is a growing feature
prior to frontier model release,albeit with notable exceptions. (Anil et
al 2023, xAI 2024) There have also been broader mandates proposed for
auditing AI systems, (Koshiyama et al 2021) though a
principles-to-practice gap exists, especially noted for AI ethics,
(Tidjoh and Khomh 2023) which is even more of a problem for the newer
auditing methods for safety.

In addition to these voluntary oft ignored self-audits, there has been
recent discussion of the need for mandatory external and independent
auditing, which would help address the principles-to-practice gap.
(Costanza-Chock et al 2022) These efforts are largely country-specific,
and are likely to create regulatory overlap failure modes seen in other
domains (Robb et al 2023), but there are at least attempts to make
standards. The recent U.S. Executive Order requires that National
Institute of Standards and Technology (NIST) work on ``Developing
Guidelines, Standards, and Best Practices for AI Safety and Security,''
following on from their Risk Management Framework (NIST, 2023), and this
(unfunded) mandate includes ``an initiative to create guidance and
benchmarks for evaluating and auditing AI capabilities, with a focus on
capabilities through which AI could cause harm,'' (E.O. 14110, 2023)
Similarly, the UK's new AI safety institute views ``publicly accountable
evaluations of AI systems'' to be a key part of its mission, (DSIT,
2023), and efforts in France have been proposed as well (Commission De
L'intelligence Artificielle, 2024).

For all the above-mentioned projects and initiatives, however, the focus
is on what should be done concretely in the evaluation of models
themselves.Despite limitations, it is a critical part of the ecosystem
necessary for audits to be useful and successful, but the efforts do not
address broader issues, including the need for a regime that will ensure
robust standards for a quickly evolving ecosystem and uncertain but
rapidly evolving ethical challenges, risks, and future threats. As
recent work has shown, even when limiting the domain to AI
accountability auditing, rather than the nascent and less well developed
field of AI safety audits, the focus of above efforts are irresponsibly
liimited to only model evaluations, and ``this ecosystem {[}which
includes broader ecosystem and product audits,{]} is muddled and
imprecise.'' (Birhane et al 2024)

\subsection{Existing Best
Practices}\label{existing-best-practices}

While efforts mentioned above for AI ethics are widespread, if
fragmented, frontier AI developers now believe that their models may
soon be capable of causing direct harm on a wide scale (OpenAI 2023d,
Anthropic, 2023). OpenAI's preparedness framework identifies
cybersecurity, chemical, biological, radiological, and nuclear (CBRN)
threats, and persuasion and model autonomy as categories of dangerous
capabilities that they expect from near-future models. While these
dangers do not obviate the need for addressing still often ignored
ethical issues, to the extent that these anticipated AI systems are
high-risk, they must be treated in line with the best practices for
other safety-critical settings such as aviation or the nuclear industry.

Therefore, we start by drawing on existing examples of industry best
practices for auditing standards, adapting them where relevant to the
setting of AI evaluation, and the role of a standards board. We also
discuss the need for a shift towards safety culture, not just formal
standards and practices, something that is well established in
safety-critical industries, but severely lacking in the AI Industry
(Manheim 2023a).

In aviation, for instance, safety culture audits are common, and FAA
assessments focus on employees' perception of safety
culture and their self-assessment of how important safety is to their
work. Similarly, the nuclear industry's governance
practices highlight the importance of regulatory oversight of safety
culture, rather than only individual safeguards. (IAEA, 2013) In other
safety-critical industries like manned spaceflight (Keyser, 1974) we see
management structures changed to give increased veto power to
lower-level decision-makers should they notice problems.

These governance norms can be mirrored in AI to ensure safety protocols
are embraced not just operationally, but also culturally. For example,
the financial sector provides a case study for how to effectively audit
and disclose risks, allowing external auditors comprehensive internal
access, ensuring transparency while maintaining confidentiality. Contra
Raji et al (2020)'s calls for internal audit, aligning with Raji et al.
(2022b), it is clear that a standards board for creating norms and best
practices can mirror financial audit and ensure that independent AI
audits can be conducted without requiring companies to publicly reveal
proprietary details of their business.

Lastly, the pharmaceutical industry's practice of making
explicit commitments to ongoing audits with requirements for publishing
results can guide AI auditing. The FDA, for instance, mandates regular
pharmaceutical inspections with publicly available results, ensuring
continuous compliance. This is in stark contrast to AI audits that have
been proposed or are currently performed on ad-hoc bases, without
external monitoring. Pharmaceuticals also provide another model for
prioritizing oversight and resource allocation based on risk. (Lawrence
and Woodcock, 2015) It also provides another warning about the costs and
unnecessary duplication and consumer harm from differing and
inconsistent national standards, with outsized harms to LMICs, (Moscou
and Kohler 2018), as well as regulatory capture of standards and
regulation (Carpenter 2013).

Most fundamentally, having a body with an explicit mandate to produce
auditing standards in the first place is essential, and in all the cases
previously mentioned (IAEA, FDA, FASB/IASB, and the FDA or its
international equivalents,) some independent body exists ensuring that
these standards are not merely common industry practices, but understood
standards. Doing so in a way that learns from these examples, and does
not fall prey to the various failure modes which are seen in those
domains is possible, and would be beneficial for all parties involved.

\subsection{Problems Specific to AI
Development}\label{problems-specific-to-ai-development}

There are a number of distinct problems faced by those attempting to
develop audit standards for AI ethics and safety. Before explaining the
three-pronged approach an Audit Standards Board would need to take, we
outline four challenges. Each of these, we feel, is poorly addressed by
the current assumption that standards development will be helpful in
limiting either speculative risks or current harms.

First, static standards created and adopted in advance of an AI system's
development are sharply limited in their ability to ensure that AI
systems are safe prior to further training or eventual deployment.
Standards produced today are unlikely to be sufficient in a year, much
less several years, and they aim at a moving and unpredictable target.
(Zhou et al, 2023) AI system development, testing, and deployment is an
iterative process that can involve a wide variety of changes. For
example, ChatGPT moved from GPT 3.5, to GPT-4, to GPT-4V, to include
Dall-E 3 integration, completely changing first the underlying models,
then the modes of interaction, all in less than a year. The process can
also change access types, from UI access to API access, to fine- tuning
and customized models. A safety or impact audit standard which would
apply to ChatGPT when released in mid-2022 would be completely
insufficient by the end of 2023.

In addition to the changes in systems over time, developing standards is
challenging due to the emergence of capabilities in foundation
models\footnote{As defined elsewhere, foundation models are large AI
  models trained on one or more modalities of data in an unsupervised or
  semi-supervised fashion that are able to generalize well to a wide
  variety of downstream tasks as they are well-suited for zero-, one-,
  or few-shot learning and for transfer learning, i.e., transferring the
  knowledge learnt on large corpora to downstream tasks via fine-tuning.
  (\hl{Bommasani et al. 2021.})}, (Bommasani et al. 2021) which suggests
that dangerous and/or difficult to anticipate or forecast capabilities
can emerge with increasingly large AI systems, when systems are scaled
up by orders of magnitude (Wei et al. 2022). The result is that both
accountability audits focused on system outcomes, and safety audit
efforts like red-teaming, are rendered obsolete. This suggests that many
proposed metrics and evaluations are likely to end up insufficient or
misleading when applied to increasingly large systems. (Corso et al.
2023) One example is the emergence of language modeling in diffusion
models, which can generate increasingly valid text inside of generated
images, which may make audit questions previously limited to LLMs, such
as biased language, necessary in image generation models as well.
Furthermore, while increased model size often enables greater
capabilities, it is not determinative, so even when we know what
possible capabilities we are concerned about, the specific capabilities
to emerge are not obvious in advance.

A third issue is that evaluations of capabilities may fall short if they
fail to capture new modes of human interaction with the models,
especially due to emerging misuse opportunities, or via extensions of
models into new systems (e.g., AutoGPT). At a minimum, extensive red
teaming is necessary, even if automated, (Zhu et al, 2023) but the risks
associated with increasingly complex and capable models will also
require novel strategies of risk assessment. There are currently
significant unsolved issues with model adversarial robustness, and it is
also possible to use highly capable models to adversarially attack each
other, making them yet more vulnerable. (Shah et al, 2023) There are
also new threats, such as the recently identified ``Sleeper Agents,''
which have emerged. (Hubinger et al 2024)

A fourth reason that standard requirements can fall short is that models
are deployed in rapidly changing technological contexts. A model capable
of programming that poses no risk of misuse for bioterrorism may later
pose risk as biological design tools become more capable. (Sandbrinck
2023, Mouton et al. 2023) Similarly, new applications and uses of the
model itself can create danger. Foundation models that only output text
were originally assumed to be incapable of multi-stage projects and
agent-like behavior. The assumption was correct, until the early 2023
creation of AutoGPT (and similar programs) that harnessed multiple
instances of an LLM to pursue multi-step plans. This risk was
speculative, and for that reason no internal audit of the model
contemplating current uses foresaw or tested for this risk in advance.
For example, even OpenAIs very flexible evaluations, which included
continuous monitoring, did not suggest the emergence of agentic
applications for GPT-3.5 in advance of their appearance.

Therefore, to achieve the intended evaluation and mitigation of risks and harms, we need not just a standards board, but one which embraces a number of practices and standards. This approach needs to do more than just evaluate and monitor the models which are produced. We split the identified needs into three areas. First, we need to audit the process, not just the product. Second, we need to change the culture of safety complacency. And finally, we need to empower auditing standards boards, not create standard auditing procedures. Figure 1 depicts these fundamental principles and
practices that should be adhered for each of these three elements of our proposed approach.

\begin{figure}[h]
\includegraphics[width=128.7mm,height=145.5mm]{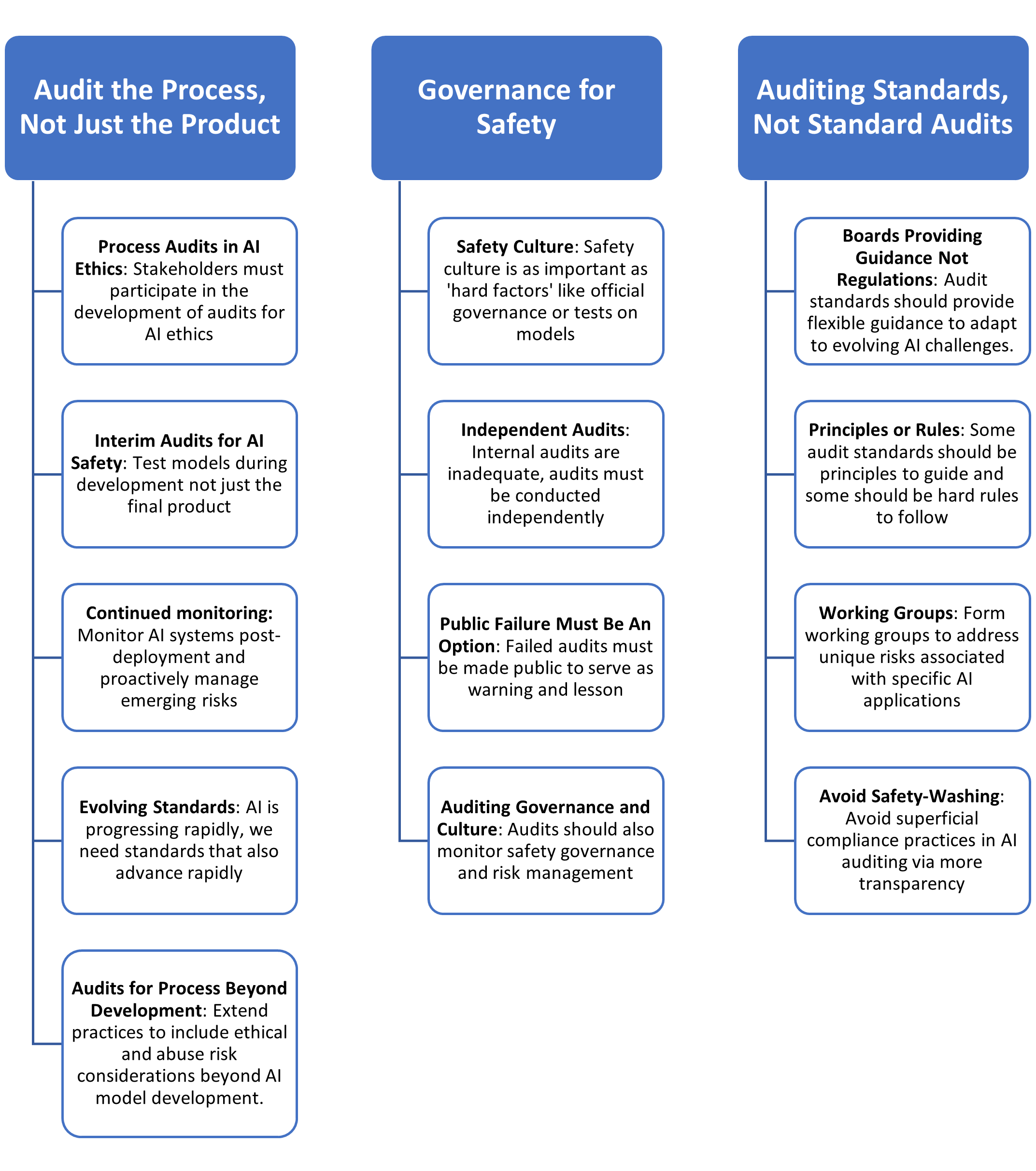}

\textbf{Figure 1:} We highlight the three elements of the approach we suggest that a standards board should take, and further identify best practices relevant to each that we recommend be promoted in the approach of boards, or other equivalent standards organizations.
\end{figure}
\section{Audit the process, not just the product}\label{audit-the-process-not-just-the-product}

As explained, a language model or other AI system poses risks that
change over time. Furthermore, black-box systems also cannot necessarily
be evaluated only as a final product, for instance, because training
data attacks can be imperceptible in the final model (Khalid et al.
2019) making evaluation of inputs and processes necessary. We see a
parallel between process evaluations for AI and cases such as
pharmaceuticals and nuclear power, where the entire design process must
be overseen. In all of these industries, it is impossible to tell just
from looking at the final result whether it is safe and effective.

There are also risks and issues that are directly about inputs or
training, such as ethical audits of data sources and human curation,
(Miceli \& Posada, 2022) and limitations and issues with RLHF for which
audit standards have already been discussed (Casper et al, 2023).
Moreover, an AI system is not a single static artifact that can be
evaluated in isolation. The increasingly interactive use of multiple
models within AI systems also means that there are complex relationships
between individual changes and ethical and safety concerns. An audit
that addresses these concerns must be broader than the evaluation of a
single product.

There are also specific model evaluation techniques that require
integration with the process. For example, use of predictions to
evaluate model capabilities, and judging the safety of a model in
relation to whether it does more than expected (or less) requires a
clear process for pre-training statements of predictions, and
post-training evaluation. This could also address problems raised by
Raji et al. (2022a), where AI projects fail to correctly do what they
were designed to do.

\subsection{Process Audits and Stakeholder
Participation}\label{process-audits-and-stakeholder-participation}

Model auditing involves and impacts a number of stakeholders, including
model developers, auditors and safety evaluators, government and
regulators, and the public. Ensuring all parties' views and needs are
represented in the development of audit processes requires not just
consultation with all parties, but continuing engagement. (Anderljung
2023b)

Raji et al. (2020) suggest viewing audits of machine learning models as
a multistage process that in part is focused on generating artifacts
which can provide insight to external stakeholders. This nicely
highlights both the need for process, which we discuss below, and the
way that the needs of stakeholders are integral to what an audit
accomplishes. While that work is focused on a subset of the concerns we
discuss, the suggestions are critical, and points to exactly the sort of
stakeholder engagement we argue is more broadly valuable.

That work, and the much broader literature, focuses on current harms
that exist from building and using the systems as intended. These
ethical issues have been highlighted by Mitchell et al (2019). To
address these harms, there is a nascent set of identified dimensions
(Krijger et al 2023) and best practices for ethics in AI --- but similar
to other concerns discussed extensively in the Fairness and Transparency
literature, this requires more than auditing the model which is created.
For example, addressing harms that emerge from biased training data or
appropriation of copyrighted data requires engagement with the data
collection and training process, rather than just the final product.
Such data audits are defined, but often remain outside of the scope of
model audits. (Birhane et al 2024)

Given the current state of practice and how far short of long-standing
and well understood standards for transparency almost all AI models
fall, (Liang et al. 2023) these recommendations must be adopted, and
failures to do so must be highlighted. But the different equities
involved mean that adopting recommendations is not enough. Instead, both
those benefiting from the model and those harmed by it need to be
represented when discussing what should be audited during these
processes.

Furthermore, systemic and structural risks, and risks from misuse,
require a sociotechnical approach to both AI safety (Lazar and Nelson
2023) and model auditing (Weidinger et al. 2023). Those harmed by
algorithmic injustice are rarely represented in current efforts towards
standards development, and the types of effort needed to include these
stakeholders (Lombe and Sherraden 2013, Bhaumik et al 2015) seems
largely absent in AI Audit standards development, and in the research
needed to enable it, respectively. (Birhane et al 2024)

In addition, as of now, it is unclear where the responsibility lies for
harms caused in ways that are not directly dependent on AI systems'
capabilities or accidents resulting from them; the interaction between
the AI system, the deployment, and the use of the systems requires
ongoing processes to understand risk. Having multi-stakeholder
engagement in defining standards and requirements benefits both society
as a whole by reducing risk, but it is also in the self interest of AI
developers as it ensures that they have clearly defined
responsibilities.

\subsection{Audits for Process Beyond
Development}\label{audits-for-process-beyond-development}

The inclusion of stakeholders in standards for auditing is critical, but
the audits themselves must be focused on more than just the models. As
Raji et al. (2020) argued, auditing needs to be present throughout the
development of a model. However, auditing for harms needs to extend
beyond the development of a model. (Raji et al 2022a)

Process audits can and should start with evaluating the company, its
structure and governance, and its safety and ethical orientation and
commitments, as discussed below, and must extend from the process of
data collection and training through post-deployment monitoring. These
require both ongoing checks, and revisions of what practices and norms
should be monitored. For example, we can consider abuse risks, which are
already occurring, that need to be addressed. Checking whether models
can be abused is critical but it is probably impossible to remove all
potential for abuse Therefore, ongoing monitoring and processes for
auditing that work will be critical.

The abuses to address will not only be those that are occurring already.
For example, mass spear-phishing is an existing threat which can
leverage LLMs to further reduce costs, (Hazell, 2023) but waiting for
proof that this type of abuse is happening before including the risk in
evaluations turns risk evaluation into a reactive and ultimately
ineffective process. And despite sensationalist presentations of
capabilities, it is clear that threats like AI-enabled bioterrorism are
currently implausible, (Mouton, 2023) but capabilities may continue to
progress. For that reason, it seems likely that by the time it is clear
such uses are possible, mitigations need to already be in place. Any
process for audit that attempts to address any of these risks will need
to move beyond point-in-time evaluation and passing or failing an audit,
and require processes for ongoing risk analysis that include routinely
revisiting questions about forms of abuse and the safeguards needed.

Societal impacts are hard to determine a priori, and will change as
these systems develop new capabilities, as society adapts to them, and
as the digital landscape changes over time. The set of concerns that
should be highlighted in audits again argues for a diverse set of
stakeholders, but also requires broader and continuing engagement and
public discussion of how and what audits must cover, rather than
one-time consultations typical of regulation or standards development.
This shows that, while sociotechnical approaches and involving broader
stakeholder groups are critical, (Lazar and Nelson, 2023) those
approaches need far more than static evaluation to be meaningfully
effective in reducing harm. As we argue, inclusive boards are a solution
that incorporates the advantages and which can mitigate many of the
problems.

\subsection{Interim Audits during AI
Training}\label{interim-audits-during-ai-training}

In addition to the ethical harms and risks of abuse of models for
current and anticipated risks, there are also risks arising from
capabilities that AIs may possess in the near future. Several groups
have noted that auditing model checkpoints and audits during training
may be necessary, both because the AI's capabilities may advance
suddenly during training, resulting in the need for new safeguards to be
developed or applied, and because auditing the non-final version (Avin
2023, Anderljung 2023, Sharkey et al. 2023) can reveal or help predict
latent capabilities.For example, safety measures like RLHF, which are
essential, can nonetheless be reversed via fine-tuning. (Gade et al.
2023, Lerman et al. 2023) This implies that safety audits of models
before fine-tuning are likely to be necessary for models which will be
made available to fine-tune.

Not only is it important to have process audits to ensure the end
product is safe, but there are also a number of mitigations for bias and
similar problems that can be applied throughout the model training
process. (Bellamy et al. 2018) For these and other reasons, we expect
defense in depth and lifecycle threat analysis to be very valuable. (Ee
2023). If it proves useful for external auditors to audit AI systems
over multiple steps during training, and throughout the process, then
they will need in-depth engagement throughout the development process.

\subsection{Rapid Progress Requires Evolving
Standards}\label{rapid-progress-requires-evolving-standards}

In various domains, best practices for risk evaluation have evolved to
recognize the need for proactive or dynamic risk evaluations. (Rasmussen
\& Suedung 2000) This means that even in much less dynamic fields than
AI, proactive engagement with risk management is vital. In artificial
intelligence, the dizzying pace of advances makes this even more
critical. But changing rules can easily create incoherent or conflicting
standards, and different standards suggestions have already created a
difficult to follow patchwork of best practices, requirements, and an
emerging regulatory tangle of overlapping and burdensome requirements.
And because standards bodies do not exist, the suggested best practices
and requirements are not necessarily updated, and there is no group that
can decide when or if obsolete regulations have been superseded.

Not only is this messy, but static standards can have perverse effects.
Specific requirements and maximums imposed by regulation often become a
target when they relate to a system property of interest, per Shorrock's
Law of Limits. (Shorrock, 2019) In the case of AI, a requirement that
systems not score over some limit on a risk scale, or not have a racial
or other bias that exceeds some numerical value, would require checking
of compliance, and if there is any expense associated with compliance,
this produces implicit pressure to get as close as possible to that
limit without exceeding it.

Another risk of static standards for evaluation of models is leakage,
(Kaufman et. al. 2011) where data about the test set, in this case, the
analyses or tests which are part of the standard, are included in the
training data. For AI systems which can generalize, this seems likely to
be particularly challenging; even if a specific question is absent from
training data, thematically similar ones could be present, allowing the
model to ``study for the test'' or the equivalent. For systems that are
under continual development, which are retrained or fine-tuned on recent
public data, any public knowledge about the contents of a test could
compromise the integrity of the test, at least to some extent.

Iterative standards development needs to occur in parallel to, and must
be informed by (but separate from), model development. That is, auditing
should not be an internal process to enable development, as discussed
below. Instead, standards setting groups should define tests which must
be passed by models if and when new concerns or capabilities emerge.
Alternatively, the emergence of unexpected capabilities may itself be a
reason to halt development or release of a model.

\subsection{Continued Monitoring and Ongoing
Analysis}\label{continued-monitoring-and-ongoing-analysis}

As well as monitoring throughout the development lifecycle, it is also
necessary to monitor systems post-deployment. This approach is already
being adopted, for example OpenAI's preparedness framework, where the
model provider monitors usage and identifies abuse or misuse of various
types. This process, however, is not currently done in a principled or
public way --- there are no clear commitments to report on these ongoing
evaluations on a regular basis or to include a specific set of
information in them. There is also no architecture for bug bounties or
patch notes. The current process is also reactive, whereas audits are
potentially and ideally proactive. Ideally, ongoing monitoring would be
part of a principled and public process that tries to anticipate future
misuse rather than reacting as new implicit capabilities are discovered.

Additionally, the OpenAI framework emphasizes forecasting future risks
to develop adequate safety and security measures ahead of
time. For example, they
intend to privately fine-tune models towards dangerous applications to
determine how far their dangerous capabilities can be stretched, and
continuously monitor their deployed models in worst case settings. This
framework's inclusion of continuously monitoring capabilities along
various dimensions could address many of the requirements we have
outlined for AI auditing -- though this auditing is done in-house, with
no reporting requirements, and no longer term commitment to the process,
so it obviously fails the goal of providing public or regulatory
assurance.

\subsection{Inadequate audits can cause
failures}\label{inadequate-audits-can-cause-failures}

Major AI developers expect that work at the AI frontier will
increasingly focus on designing AI systems to autonomously pursue
large-scale goals (Anthropic 2023, OpenAI 2023, Christiano et al. 2023).
If such systems are developed, then audits must also ensure both that
these autonomous agents do not have discriminatory or otherwise
unethical consequences, and that their goals are and remain more broadly
aligned with human values and ethics over time. However, internal teams
repeatedly testing these sophisticated AIs against simple audits could
influence them to develop unintended and dangerously deceptive
behaviors. This could directly cause dangerous failures, or at least
lead to complacency about safety as dangers are merely papered over by
trained success in audits.

Christiano identifies three specific dynamics that make it hard to
detect misalignment in AI systems. First, the "simple training game" is,
where an AI behaves sycophantically to make deception more difficult to
notice. Second, "deceptive alignment" is where AI has hidden goals other
than what is apparent in its performance and predicts what will fulfill
these goals over a long timescale. Both have already been seen to some
extent in the wild. (Park et al, 2023) Third, "gradient hacking" is,
where AI prevents training from changing its hidden long-term goals
(Christiano et al. 2023). It is unclear the extent to which any of these
will occur, but there is at least a strong reason to be concerned that
poorly managed audits will create these risks. Of course, auditing
versions of models without alignment or safety mitigations implemented
-- as suggested in the OpenAI Preparedness Framework (OpenAI 2023) -- is
one crucial way of addressing this risk, as is ensuring that audits are
not part of the internal development process.

\section{3 Governance for safety}\label{governance-for-safety}

Building on the work of Raji et al. (2020), we note that culture can be
studied at multiple levels. The nuclear industry (IAEA 2013) draws on
Schein (1992) to look at three levels of culture, artifacts, as
discussed by Raji, espoused beliefs and values, and underlying
assumptions. As the nuclear industry has found, technical safeguards are
insufficient, and building a culture of safety is critical in minimizing
failures. The airline industry, among others, has found the same. (CANSO
2008) A number of concrete suggestions for how this impacts audits more
specifically will be highlighted below, but the overarching theme is
that change in the industry requires industry buy-in to standards and
norms, of the type facilitated by standards boards in which they have an
active role, but undermined by proliferating standards.

However, as pointed out by Manheim, (2023a) there are many challenges in
building a culture of safety in the AI industry. In the context of
auditing, these challenges are critical. There are two aspects to
addressing safety culture that are relevant to audits; the way that
safety culture makes audits more effective, and the way that audits can
ensure safety culture. For both, formulaic and fixed standards seem
insufficient. Until recently, the AI industry was not focused on these
issues and is complacent as regards safety.

\subsection{Safety Culture Enables Effective
Audits}\label{safety-culture-enables-effective-audits}

Auditing systems is not, in and of itself, a method to reduce risks.
Instead, it functions as part of an ecosystem of regulation, risk
analysis, internal controls, public oversight, and changes to the
systems themselves in response to all of these factors. Because of this,
the value of auditing in improving safety lies partly in how those being
audited view the process and anticipate or respond. In domains where
audit is adversarial or viewed as superfluous, knowledge of problems
does not lead to addressing them; if any attention is paid, it can lead
to deception in order to move forward, or at best patches that minimally
address the deficiencies. Designing metrics that can be evaluated in
such an environment is possible, but very challenging - and a key
strategy for overcoming the challenge is adaptability and flexibility.
(Manheim, 2023b) It seems likely that standards boards would make this
type of adaptability and flexibility easier.

In contrast to such antagonistic relationships between auditing and
business, there are many domains where auditing is aligned with the
mission and values of those audited. In safety-critical industries like
the nuclear industry or aviation, safety of the power plants or
airplanes is vital to the overall success of an organization, and safety
failures are business failures. These are industries where safety is
valued, and firms take note of highlighted deficiencies, which can then
be integrated into plans for improvement. If it is important for the AI
industry to emulate the positive examples, a critical role of those
tasked with monitoring and evaluating safety of AI systems is promoting
cultural change in the industry. Unless and until safety is a goal of
those producing the models, among other things, auditing will lead to
adversarial engagement, or at best grudging compliance. Such cultural
change also benefits greatly from collaborative engagement in standards
settings.

\subsection{Internal Auditing is
Insufficient}\label{internal-auditing-is-insufficient}

Internal auditing is one way to try to build alignment between auditing
and development. This has a number of advantages over external audits,
as clearly discussed by Raji et al. (2020) However, public assurance
requires more than just internal auditing. Whether audits are conducted
internally, externally, or (preferably) both, it is essential to
establish and enforce audit reporting standards independent of the firms
being audited. Among other concerns, this is because internal auditing
without broad input about the concerns that must be addressed can easily
become safety-washing. (Lazar and Nelson, 2023) A business which
declined to have their financial records audited externally because they
did sufficient checking internally would be farcical, and a business
which decided to set the standards which it used would be similarly
unacceptable.

For this reason, attempts to bring auditing under the control of firms
developing models, such as OpenAI's push to keep their red-teaming under
their own control, (OpenAI 2023b) is a worrying development. Similarly,
the ``Frontier Model Forum'' (OpenAI 2023c), while laudable as an
attempt to ensure that standards exist, could allow industry to maintain
control over what those standards will be.

At the same time, developers' concerns should also be a critical
consideration. Developers claim that proprietary concerns merit keeping
capabilities evaluations in house, and some go so far as to suggest that
these concerns merit exclusion of third-party evaluators (Weidinger et
al. 2023). As noted, this is fundamentally opposed to trusted audits,
but the risk is critical. Standards setting bodies must find ways to
balance external inputs and requirements with internal control, as has
been done by standards boards and regulators in other industries.

For example, model auditing processes, and the standards that inform
them, must be careful to take stringent steps to protect
developers' intellectual property. Toward these ends,
it is not unreasonable for third party evaluators to be embedded in
organizations and subject to confidentiality agreements---it can work to
the advantage of auditors to have unfettered access to engineers and
scientists developing the systems, and without such access, evaluations
might be far less effective. Despite that, contrary to the status quo,
the confidentiality agreements must themselves be public, to allow
consumers of these independent audits to know what the auditors may be
prevented from saying. Negotiations between model auditing providers and
firms, however, make it difficult for auditors to insist on such terms.

Only external standards that balance these and similar considerations,
developed openly as part of an independent and iterative process with
industry participation, have been shown in other comparable cases to
lead to reasonable assurances of safety. In such a broad-based process,
those who are not dependent on firms, including regulators and the
public, must push for more openness, where auditors would not be willing
or able to do so. A forum which sets standards is an ideal setting for
this.

\subsection{Audits Should Provide Transparency, Not Create
Requirements}\label{audits-should-provide-transparency-not-create-requirements}

One critical component of an audit is to create common knowledge of what
is and is not true, and what has or has not been done. While it is
important for standards for audits to be created, that does not mean an
audit itself implements or enforces a standard. For this reason, an
audit may show that a system is capable of a certain class of abuse, or
is not, and in either case, that is an acceptable audit outcome, even
when it is a substantive failure. Similarly, a firm which is audited
financially and is found to have debts in excess of assets may be
insolvent, but the audit was not failed. The equivalent determination
for an AI system or company would be that it abides by a given standard
or follows a regulatory requirement, or not.

For example, there are debates about the legality or morality of
training models on copyrighted material. While an audit standard will
not resolve this debate, an audit can determine whether such material
was included in the training data, what efforts were made to exclude
both copyrighted material, and material which creators explicitly
request to have excluded. It will then be up to regulators and the
public to insist on reasonable rules. Similarly, audits of AI models
should detail whether internal or external red teaming processes
occurred, and what they did or did not cover. This is different than a
requirement to do so - and whether such a requirement should exist is
again, up to regulators and lawmakers, not those who design or carry out
audits.

\subsection{Audits Must Allow for
Failure}\label{audits-must-allow-for-failure}

Despite the above points, if an audit cannot be failed, it cannot serve
its full purpose as an audit. In domains where auditing does function as
an independent check on firms, such as in financial accounting, the
auditors will collect instances of material misrepresentation and report
them, rather than simply demanding compliance. This is because such
post-hoc compliance is ineffective at addressing core concerns; a firm
which lies to auditors should not have their remaining claims trusted.
For this reason, requiring them to correct the specific instances in
which the misstatements occurred does not change the need to report that
misrepresentation occurred.

Similarly, a firm which builds unsafe systems for audit, expecting that
the unsafe system will be remedied during the audit, has already failed
to build a safe system. For this reason, allowing them to correct the
specific instances in which the model is dangerous is insufficient. Not
only is it unlikely to fully address the problems uncovered, but the
safety failures should be reported by auditors even when the firm
addressed the failures before release. This means that in practice,
auditors should expect to see the reports for safety failures that were
caught internally by a responsible company's monitoring and reporting,
not by auditing - and the absence of such events being flagged before
auditing should be seen as a red flag, not a positive sign.

Unfortunately, a variation on this can easily be used perversely to
render auditing far less valuable. Audits that are within the
development cycle are often used as a way to iterate on failure rather
than as a check. In such cases, the supposed existence of an audit is
(perhaps unintentionally) actually enabling development rather than
acting as a check. This is especially acute if the audit is under the
control of those being audited, either because the audit is performed
internally, or with nondisclosure agreements. In fact, this is
explicitly the process that OpenAI's Preparedness framework adopts,
where high or critical risk systems are not stopped, and instead,
mitigations are put in place so that the systems can be argued to be
below the internally defined threshold. (OpenAI 2023d) For this reason,
audits should be independent of development, and should have the mandate
to honestly report what was found, acknowledging failures rather than
ensuring that failures are individually addressed.

Even for less serious concerns, and for audit requirements that do not
impose a standard at all, as discussed above, allowing firms to address
problems without announcing the issues or changes prompted by the audit
is a failure. For example, models that are fine-tuned to eliminate
individual problems still have capabilities that can be abused by prompt
engineering, or by fine-tuning to reverse the new safety measures.
Unfortunately, quietly and iteratively fine-tuning failures away as they
are discovered, either before or after release, seems to be the current
modus operandi. (Mitchell, 2019) But post-hoc patches do not provide
public knowledge of fallibility and cannot serve the same purpose as
auditing. For parallel reasons, audits of training data that identify
biases, or problematic content of various types, must report this fact
even if the training data is amended to exclude that content.

Non-public failures and responses also fail to alert others about
measures which individual firms found to be effective. This is true of a
broad variety of harm mitigation measures, and while businesses
developing AI have reasonable incentives not to publicize proprietary
methods, if the culture of proprietary technology applies to safety and
ethics, insights about how to address problems will remain unknown, and
safety of other systems will be impeded. In this case, the benefits of
openness parallel early claims for the necessity of open-source in
machine learning endorsed by both Bengio and LeCun, (Sonnenburg 2007)
rather than their recent debates over whether model weights should be
open-source. (Nuñez, 2023) At the same time, auditors will presumably
have the prerogative not to publicize measures which would be
compromised by public knowledge.

\subsection{Auditing for Governance and
Culture}\label{auditing-for-governance-and-culture}

To conclude the discussion of the relationship between auditing and
building a safety culture, we reiterate our earlier point that a
sufficient audit must have a scope that includes the firm and its
governance, management, and culture. Ideally, auditing should inform
everything from strategic goals to monitoring. (Filyppova et al., 2019)
Once that is the case, auditing should be used to understand and monitor
the safety culture.

For example, audits should include a review of the risk management
systems internal to the firms being audited. A company with a healthy
safety culture would have well documented safeguards for internal
reports of safety failures and incentives for reporting risks and
failures, would use risk registers to document what risks they address
and which have been raised, would have robust systems in place to
address each of the risks, would have systems for raising and addressing
internal safety concerns, and would have clear procedures for stopping
development or training if concerns are raised. All of these systems
would have clear processes in which actions taken are documented, and in
each case, an audit would create artefacts, as discussed by Raji et al.
(2020) documenting how the systems have been used, the concerns and
risks they currently address, and exercises done to test the system's
ability to respond.

Surveys of employees' view of risks and their understanding of how the
company engages in proactive identification of new sources of risk are
another type of tool that auditors can and should use. Beyond this,
however, firms can also use similar tools internally to proactively
assess and reinforce safety culture. Similarly, process improvement
targeted at improving risk management can be a result of audit
recommendations, but far better than needing auditor feedback is
auditors ensuring that the firm itself has internal process improvement
methods that focus on safety and identify risks and concerns, and
address them openly, in ways that auditors can observe. To ensure this,
it seems likely that auditors should, as a part of their audit, ensure
that management and oversight systems are in place for reporting
failures, either publicly, or at least to the auditors. All of these
suggestions are issues that an audit standards board should discuss and
choose how auditors should implement.

\section{4 Auditing standards body, not standard
audits}\label{auditing-standards-body-not-standard-audits}

We do not have sufficient standards for software safety in general, nor
for ethical usage of such systems, compounding the problems with making
AI systems safe. Building such standards is important, but disparate
proposals and a lack of unification of standards undermines
accountability. Industry standards are a critical step in moving
forward, but allowing one-time processes to develop standards to
dominate the conversation is undermining the effectiveness of these
efforts. As noted initially, in comparable industries, there are bodies
explicitly tasked with ongoing development of standards, and because of
the likely importance of AI in industry in coming years, no less should
be expected here.

\subsection{Boards providing guidance, not
regulations}\label{boards-providing-guidance-not-regulations}

Best practices from other fields is that standards develop as risks and
practices change, but this process is ongoing, so that standards are
constantly evolving. More than that, while technical research informs
audit boards like FASB for accounting, or regulators like OSHA for
safety, these standards are part of an ecosystem, not an endpoint. This
is uncontroversial in other domains; FASB is asked to provide guidance
on a routine basis whenever new accounting issues arise, and OSHA
provides interpretation letters to clarify how requirements apply in new
cases.

We have also seen how this fails in other domains. Organizations like
NIST provide important standards for cryptography, updated routinely,
and provide technical reports on blockchain systems, but regulation and
standards cannot keep up with actual applications in digital securities.
This happens because the faster the systems develop and change, the more
agile the standards must be. This makes flexibility critical, and
bolsters the case for independent boards rather than direct government
regulation. It also highlights the importance of guidelines for auditors
that can be interpreted into specific audit standards, rather than
standards bodies constantly changing requirements.

In parallel cases, government regulators have required corporations to
follow guidelines developed privately. For example, ISO standards, which
are developed by experts from a variety of national standards bodies,
business groups, and others, are not themselves legal standards.
Regulators must often adapt or adopt the standards, and the standards
are often enforced in contracts. This often helps create a norm of
compliance with an evolving standard.

\subsection{Balancing Principles and
Rules}\label{balancing-principles-and-rules}

Whichever class of oversight and development of audit standards occurs,
one critical question is whether to use principles based standards, or
rule based standards. This parallels but does not exactly match the
question of static versus adaptable standards; there can be unhelpfully
static principle based standards, and dynamic and rapidly adapted rule
based standards, but these are harder than having good principles that
give rise to adaptability. And principles based standards have been
shown to lead to better auditing than rule base standards in other
important ways. (Peytcheva 2014, Schipper 2003)

Principle based standards also provide much more adaptability for
rulemakers to provide useful guidance. For example, in the above
discussion differentiating between requirements and standards, it was
noted that transparency about practices can be a viable alternative to
mandating specific practices. In this case, transparency should be a
principle, not a specific requirement to announce that some specific
method was (or was not) used. Otherwise, firms will use standard
practices to fulfill requirements and ensure that the audit reports they
did so, making the audits far less flexible and more of a routine. The
alternative instead provides flexibility to do more or use newer and
more powerful methods for assuring some specific objective is achieved,
while ensuring that auditors and consumers of the audit can still
understand if the methods used fall short in some regard.

But as is often the case for binary distinctions, some combination of
the two is likely both inevitable, and wise. Principles relevant in this
context would perhaps include transparency about model training and
inputs, preregistration of training details including safety guards,
prior commitments to which safety measures are employed during each
phase of release, commitments for ongoing monitoring and plans for
mitigating negative impacts, and external verification of the above.
Rules setting minimum standards for these are then needed, but these
will inevitably be both incomplete and unreasonably restrictive and will
likely require adaptation of the rules in individual cases via guidance
from the standards board.

\subsection{Working Groups for Applications-Specific
Risks}\label{working-groups-for-applications-specific-risks}

Regulations which exist in many specific domains often require intention
or an act by a human, and it is unclear how this applies, or whether
prosecution of violations by AI systems would be viable. There is also a
lack of expertise in these domains, and the complexity of current AI
systems means that specific standards for each domain are needed, and a
standards board would be in an excellent position to provide assistance
and guidance.

Some applications, including law enforcement and medicine, obviously
have specific risks that must be understood, and weighed against
benefits of the application before allowing them to proceed. Domains
with regulations restricting human practice such as engineering or
medicine already have standards, but the groups setting and updating
those standards are both unaware of the AI-specific risks, and not
currently able to regulate AI systems. The current approach seems to be
that AI systems are either narrow, or are trained to refuse to answer
within domains where issues exist. This is a reasonable stopgap measure,
but unlikely to be sufficient moving forward.

For this reason, application audits are needed. These are necessary to
ensure ethical and safe usage of AI systems or algorithms which may
themselves be safe. (Akula and Garibay, 2021) These application-specific
audits will involve components required of both the system developers,
and the application developer, so general standards will need to
anticipate the requirements of applications. Audits of the general model
will need to provide guidance for what is acceptable for more specific
applications. For example, a generative AI for producing CAD or similar
outputs must be held to a higher standard if it will be used for car
parts, much less dental implants, than if it is used to generate content
for online games.

Of course, AI systems which are intended to be generally capable pose
additional challenges. User agreements which list standard disclaimers
about not relying on a model for medical, legal, or other advice, and
fine-tuning to make AI systems disclaim their answers may or may not be
legally relevant, but will have limited effect on how these systems are
actually used. To the extent that a model can be urged to provide a
service, there is an ethical if not legal requirement to ensure it does
so responsibly. And either because it may be used in specific
applications, or because it may be used directly in these higher risk
domains, such general models may be required to pass any and all domain
specific requirements.

Whether considering general or narrower AI systems, it is easier and
more useful to consider holistic approaches and expert groups able to
respond dynamically to changing needs, rather than evaluating and
addressing each class of harm with different sets of incomplete and
overlapping audits. To enable this, coordination between regulators,
industry, auditors, domain experts, ethicists, and the public will be
essential. Flexible institutions able to coordinate with all of these
groups, rather than a narrow standards process, seem well suited to the
challenges - both those which are anticipated, and those which will
inevitably arise in the future.

\subsection{Safety-washing and industry
dynamics}\label{safety-washing-and-industry-dynamics}

An incentive structure for AI auditing needs to be utilized that ensures
not only high fidelity audits of frontier models, but high rigorous and
robust audits at a more granular level, specific to different classes of
risks, as evaluating and red teaming different classes of risks may
involve quite different and unique skill sets and experience. (Frase
2023) Despite the many unique challenges, the concerns mentioned
parallel those in other industries, and therefore share some failure
modes.

The default for many modern industries is some degree of regulatory
capture, competition creating races to the minimum standards, asymmetric
assignment of responsibility to avoid culpability, and audits that
provide more reputational to substantive benefits. We see this
everywhere from chocolate production and modern slavery
(Gutierrez-Huerter et al, 2023) to mining (Gold et al 2015, Nygren
2018), and expect similar dynamics to continue to frustrate efforts in
AI safety (Guest et al 2023) and in AI ethics (van Maanen, 2022).

Current efforts to build standards will fall prey to both the appearance
and reality of corporate pressures towards failure. Given the widespread
damage caused by current use of algorithmic decisionmaking, (O'Neil
2017) and the rapid and ongoing near-universal adoption of AI systems
with even less transparency and more critical tasks, industry capture
would be disastrous. The challenge of regulation is worsened by the rush
towards investments and focus on profits, which will ultimately hurt the
industry. Failures, even reasonable ones, would lead to public pushback
against even beneficial AI, and such failures are inevitable, especially
if standards are dictated by narrow interests.

It makes sense, then, that there are also efforts underway to preempt
regulation, and not just via lobbying. For example, efforts like
``Responsible AI Maturity Models'' which are promoted by Microsoft
(Vorvoreanu et al, 2023) and industry-led groups such as Responsible.AI
(2024) are nearly explicit attempts to preempt regulation by creating
voluntary industry standards. However, its inability to serve the
necessary role is obvious given the concept of a maturity model --
common in unregulated industries and those where standards do not exist
or are not enforced. In addition, we view the narrowness of the scope of
what responsibility means, and the non-participatory nature of their
development, which excluded regulators and the public, as clear
indications of their purpose.

The world has seen both more and less successful approaches for building
safety and ethical standards. Given the expected importance of AI in the
industry, it is in the enlightened self-interest of industry to ensure
that robust, open, and neutral auditing standards are developed. And
given the concern about the criticality of auditing frontier models
against larger future risks,, it is imperative that auditing's rigor is
prioritized with redundancy and robust standards and certification
procedures for auditors.

\section{Conclusion}\label{conclusion}

An audit standards board would address many issues with AI audits
highlighted in the paper which seem problematic, and efforts in this
direction seem valuable. As noted initially, standards boards for safety
and similar audits are common across many industries, in various forms.
Despite the many differences between AI and those in other fields, the
essential reason for needing such a group is unchanged; audits must be
fair, trusted, and fit to the purpose of enabling stakeholders to make
determinations about the facts. None of those goals can be accomplished
without having the standards set independently. A proliferation of
individual standards seems unlikely to accomplish this goal, whereas a
board of people actively building auditing systems and performing
audits, with cooperation from regulators and industry, seems far more
promising.

Such a board would not, in and of itself, solve problems with regulatory
capture, much less with the fundamentally difficult problems of building
artificial intelligence that doesn't exacerbate current harms or the
threatened future risks. Open work by such a board would, however, make
it clearer to everyone what progress was or was not occurring. To the
extent that AI governance is an information problem, (Anderljung 2023b)
a standards board can help remediate that problem. And to the extent
that ethical and safe AI can be enhanced by wider sharing of best
practices, standards, and methods, this seems like the right direction
to move.

However, the groups which need to push for such a standards body are not
the academics studying the problem, but a coalition of the companies
building these systems, the auditors performing evaluations, and the
regulators, government bodies, and international agencies who are
charged with oversight. Consensus about AI audit as one important path
forward is a promising recent development, and we are hopeful that this
next step, of having an active standards board that sets audit standards
and provides resources for performing them, also becomes a
widely-accepted next step in building safer and more fair AI systems.

\bibliographystyle{plain}
\bibliography{references}
\nocite{*}

\begin{small}
We would like to thank two of the anonymous reviewers for their feedback.
\end{small}

\end{document}